\documentclass[prd,aps,nofootinbib,12pt]{revtex4}
\usepackage{xspace}
\usepackage{mathrsfs}
\usepackage{amsmath}
\usepackage{amssymb}
\usepackage{epsfig}
\usepackage{graphicx}
\usepackage{booktabs}
\usepackage{multirow}
\usepackage{subfigure}
\usepackage{bm}
\usepackage{times}
\usepackage{braket}
\usepackage{color}
\usepackage{slashed}
\usepackage{hyperref}
\usepackage{threeparttable}
\usepackage{physics}
\allowdisplaybreaks

\definecolor{Red}{rgb}{1.,0.,0.}

\definecolor{Blue}{rgb}{0.,0.,1.}

\definecolor{nicered}{rgb}{0.7,0.1,0.1}
\definecolor{nicegreen}{rgb}{0.1,0.5,0.1}
\hypersetup{colorlinks,citecolor=nicegreen,linkcolor=nicered}

\begin{document}

\title{CP violation analysis of $D^0\to M^0 K \to M^0(\pi^\pm \ell^\mp \nu)$}

\author{Wen-Jie Song} \email{joycesong0213@gmail.com}
\author{Shi-Qi Wang}
\author{Qin Qin} \email{qqin@hust.edu.cn}
\affiliation{School of physics, Huazhong University of Science and Technology, Wuhan 430074, China}
\author{Ya Li\footnote{corresponding author}} \email{liyakelly@163.com}
\affiliation{Department of Physics, College of Sciences, Nanjing Agricultural University, Nanjing 210095, China}


\begin{abstract}
CP violation in the charm sector is highly sensitive to new physics due to its small predicted value within the standard model. By this work, we investigated the CP violation in the cascade decay process $D^0(t_1) \to \pi^0 K(t_2) \to \pi^0(\pi^\pm \ell^\mp \nu)$. Our results indicate that the CP violation induced by the interference of unmixed $D$-meson decay amplitudes dominates, with a peak value reaching $5 \times 10^{-3}$. This is one  order of magnitude larger than the sub-leading contribution, namely the double-mixing CP violation. Furthermore, the CP violation in the decay channel with the semileptonic final state $\pi^- \ell^+ \nu_\ell$ is one to two orders of magnitude larger than that in the channel with $ \pi^+ \ell^- \bar{\nu}_\ell$. We propose that the CP asymmetry of the combined two decay channels can be measured experimentally.
The resulting value is approximately half of the CP violation observed in the $\pi^- \ell^+ \nu_\ell$ channel, with the dominant contribution still reaching the order of $10^{-3}$. 
This approach offers the advantage of eliminating the need for flavor tagging of the initial $D$ meson, thereby avoiding associated efficiency losses.
\end{abstract}
\maketitle

\section{Introduction}
CP violation plays an important role in explaining the matter-antimatter asymmetry in the universe~\cite{Sakharov:1967dj} and in searching for new physics beyond the Standard Model (SM). 
The discoveries of CP violation have been made in $K$\cite{Muller:1960ph,Christenson:1964fg,KTeV:1999kad}, $B$\cite{BaBar:2001ags,Belle:2001zzw,BaBar:2004gyj,Belle:2004nch,LHCb:2013syl} and $D$\cite{LHCb:2019hro} meson decays.
Extensive research has been well established in the $K^0$ and $B^0_{(d,s)}$ systems and has been found to be consistent with the predictions of the SM.
CP violation in the charm sector is expected to be very small in the SM since it is theoretically suppressed by the Glashow-Iliopoulos-Maiani mechanism~\cite{Glashow:1970gm}, which, however, can be possibly enhanced by new physics effects.
Therefore, the study of mixing and CP violation of the $D^0$ meson can provide a unique probe of NP in flavor-changing-neutral currents in the up-type quark sector, complementary to that of $K^0$ and $B^0_{(d,s)}$ mesons.

The cascade decay $D^0\to M^0 K \to M^0(\pi^\pm \ell^\mp \nu)$, where $M^0$ is a purely neutral meson, has multiple CP violation sources. In addition to the traditional three types of CP violation, namely, CP violation in decay, in mixing, and in interference between $D$ decay and mixing, they also include the CP violation in interference between $K$ decay and mixing~\cite{Grossman:2011zk,Yu:2017oky}, and the double-mixing CP violation~\cite{Shen:2023nuw,Song:2024jjn}. Therefore, these decay channels hold the potential to observe the two novel sorts of CP violation. They also provide an ideal platform for analyzing the multiple CP violation mechanisms, {\it e.g.,} extracting the corresponding weak phases. It turns out that the magnitude of CP violation effects in these channels can reach $\mathcal{O}(0.1-1)\%$, which can be observed in current and future charm-factory experiments~\cite{BESIII:2020nme,Belle-II:2018jsg,Cerri:2018ypt,Ai:2024nmn,Charm-TauFactory:2013cnj,Achasov:2023gey}.

The $D^0-\bar D^0$ mixing mechanism can also be probed in these decay channels. The neutral charmed meson $D^0$ can oscillate to $\bar{D}^0$ via the short-distance $W^{\pm}$ exchange or long-distance hadronic rescattering effects.
The effects of oscillation or mixing can be characterized by the mixing parameters $x_D$ and $y_D$, which are defined in terms of the mass and width difference between the two neutral $D$ meson mass eigenstates, normalized to the neutral $D$ meson decay width, respectively.
The observed $D$ meson mixing~\cite{LHCb:2022gnc,LHCb:2022cak} is significantly greater than the short-distance analysis of $D^0$-$\bar{D}^0$, including next-to-leading-order QCD corrections, which yields $x_D, y_D\sim 10^{-7}$~\cite{Golowich:2005pt} by four orders of magnitude. Subsequent theoretical studies have updated the predictions for the mixing parameters to the order of $10^{-3}$~\cite{Li:2020xrz,Li:2022jxc,Jiang:2017zwr,Cheng:2024hdo}.  
The CP violation observables in the cascade $D^0\to M^0 K \to M^0(\pi^\pm \ell^\mp \nu)$ decays are sensitive to the mixing parameters $x_D$ and $y_D$, thus serving as a benchmark for comparison with experimental results.

By this work, we intend to study the CP violation of the $D^0\to M^0 K \to M^0(\pi^\pm \ell^\mp \nu)$ decay. We will take $M^0$ as $\pi^0$ as an example, and the results apply directly to $M^0=\rho^0,\omega$. This decay has multiple interference paths, induced by $D$ mixing, $K$ mixing and $D \to K^0,\bar{K}^0$ decay amplitudes, as depicted in Fig.~\ref{fig:decay}. Diverse CP violation effects arise from interferences between different evolution paths. It turns out that the CP violation induced by the interference of unmixed $D$-meson decay amplitudes dominates, including the CP violation in $K$ mixing, and the CP violation in interference between $K$ decay and mixing. The double-mixing CP violation is smaller by one order of magnitude. 
Furthermore, the CP violation in the decay channel with the semileptonic final state $\pi^- \ell^+ \nu_\ell$ is one to two orders of magnitude larger than that in the channel with $ \pi^+ \ell^- \bar{\nu}_\ell$. We propose that the experiment incorporate these two decay modes to facilitate the measurement of CP violation. A key advantage of this approach lies in the fact that it circumvents the requirement for flavour tagging of the initial $D^0$ state.

In the remainder of the paper, we will first present the relevant formulas for $D^0(t_1) \to K(t_2) \to f$. We will then focus on the numerical analysis of two specific decay channels: $D^0(t_1) \to \pi^0 K(t_2) \to \pi^0(\pi^-\ell^+\nu_\ell)$ and $D^0(t_1) \to \pi^0 K(t_2) \to \pi^0(\pi^+\ell^-\bar{\nu}_\ell)$. Possible experimental search strategies will also be discussed.

\section{Formulae}
The flavor eigenstates of the neutral meson, $\ket{M^0}$ and $\ket{\bar{M}^0}$, are distinct from its mass eigenstates, $\ket{M_{H,L}}$. These states are related by the complex coefficients $q$ and $p$~\cite{ParticleDataGroup:2024cfk}
\begin{align}
\ket{M_{H,L}} = p\ket{M^0}\mp q\ket{\bar{M}^0},
\end{align}
where $M_H$ and $M_L$ represent the heavier and lighter mass eigenstates, respectively. Assuming CPT invariance, the time evolution of the neutral meson is formulated by~\cite{ParticleDataGroup:2024cfk}
\begin{align}
&\ket{M^0(t)} = g_+(t)\ket{M^0} - \frac{q}{p}g_-(t)\ket{\bar{M}^0},\nonumber\\
&\ket{\bar{M}^0(t)} = g_+(t) \ket{\bar{M}^0} - \frac{p}{q}g_-(t) \ket{M^0},
\end{align}
with
\begin{align}
g_\pm(t) = \frac{1}{2}\Big[\exp\pqty{-im_H t - \frac{1}{2}\Gamma_H t} \pm \exp \pqty{-im_L t - \frac{1}{2}\Gamma_L t} \Big].
\end{align}
The mixing parameters $(q/p)_{D,K}$ can be formulated in terms of the complex parameters $\epsilon_{D,K}$ as
\begin{align}
\frac{q_M}{p_M} = \frac{1 - \epsilon_M}{1+\epsilon_M},
\end{align}
where ${\rm{Re}}(\epsilon_K) \approx 1.6 \times 10^{-3}$, ${\rm{Im}}(\epsilon_K) \approx 1.5 \times 10^{-3}$~\cite{ParticleDataGroup:2024cfk}, ${\rm{Re}}(\epsilon_D) \approx 1.5 \times 10^{-4}$ and ${\rm{Im}}(\epsilon_D) \approx -2.7 \times 10^{-5}$~\cite{Li:2022jxc}. 

\begin{figure}[htbp]
    \centering
    \includegraphics[keepaspectratio,width=14cm]{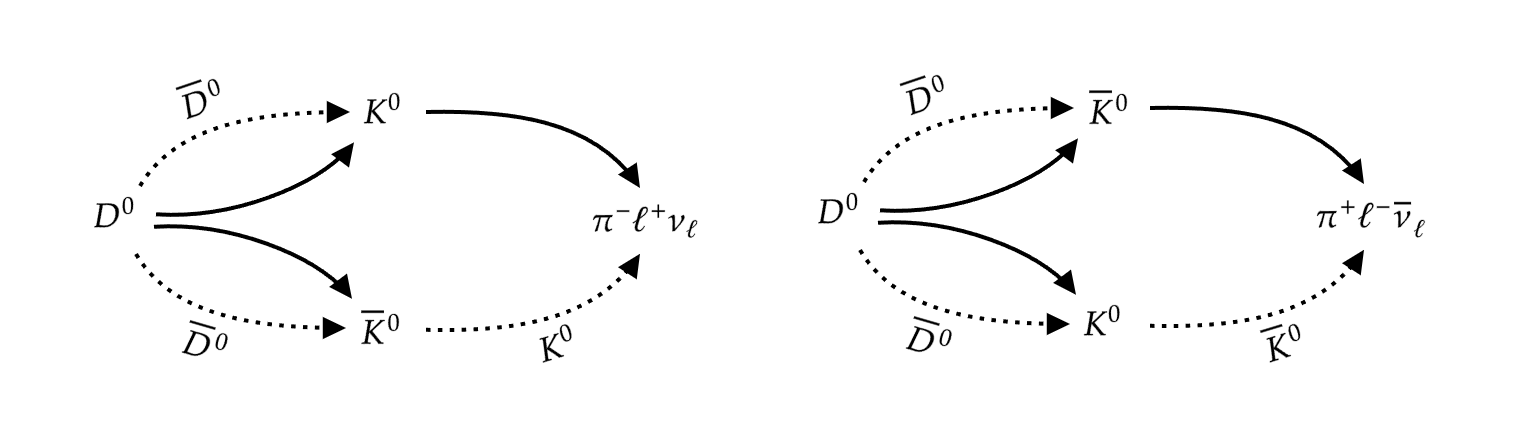}
    \hspace{0.1cm}
    \caption{The two decay modes of $D(t_1) \to K(t_2) \to \pi\ell\nu_\ell$.}\label{fig:decay}
\end{figure}

Considering the cascade decay process of neutral $D$ mesons, where a neutral $D$ meson decays into a kaon and a spectator meson, with the kaon subsequently decaying into a semi-leptonic final state $f$, as depicted in Fig. \ref{fig:decay}. A solid line and a dotted line represent the direct decay and flavor oscillation of the neutral meson before decay, respectively. Following the definitions established in~\cite{Shen:2023nuw, Song:2024jjn}, the corresponding CP asymmetry for this decay mode exhibits a two-dimensional time dependence
\begin{align}
A_{CP}(t_1,t_2) &= \frac{\Gamma[D^0(t_1) \to K(t_2) \to f] - \Gamma[\bar{D}^0(t_1) \to K(t_2) \to \bar{f}]}{\Gamma[D^0(t_1) \to K(t_2) \to f] + \Gamma[\bar{D}^0(t_1) \to K(t_2) \to \bar{f}]}.\label{ACP}
\end{align}
Here, the oscillation times $t_1$ of $D^0$ and $t_2$ of $K^0$ refer to the time durations in the rest frames of $D^0(\bar{D}^0)$ and $K^0(\bar{K}^0)$, respectively. The amplitude $\mathcal{M}(t_1,t_2)$ for the cascade decay process $D^0(t_1) \to K(t_2) \to f$ and the amplitude $\bar{\mathcal{M}}(t_1,t_2)$ for its CP-conjugate process $\bar{D}^0(t_1) \to K(t_2) \to \bar{f}$ are given by
\begin{align}
&\mathcal{M}(t_1,t_2) = \braket{f}{K^0(t_2)}\braket{K^0(0)}{D^0(t_1)} + \braket{f}{\bar{K}^0(t_2)}\braket{\bar{K}^0(0)}{D^0(t_1)},\nonumber\\
&\bar{\mathcal{M}}(t_1,t_2) = \braket{\bar{f}}{K^0(t_2)}\braket{K^0(0)}{\bar{D}^0(t_1)} + \braket{\bar{f}}{\bar{K}^0(t_2)}\braket{\bar{K}^0(0)}{\bar{D}^0(t_1)}.\label{Amplitudes}
\end{align}
To substitute Eq. \eqref{Amplitudes} into Eq. \eqref{ACP}, it is beneficial to expand the result into
\begin{align}
A_{CP}(t_1,t_2) &= \frac{C_+(t_2)\abs{g_{+,D}(t_1)}^2+C_-(t_2)\abs{g_{-,D}(t_1)}^2 + F_h(t_1) S_h(t_2) + F_n(t_1) S_n(t_2) }{C_+^\prime(t_2)\abs{g_{+,D}(t_1)}^2+C_-^\prime(t_2)\abs{g_{-,D}(t_1)}^2 + F_h(t_1) S_h^\prime(t_2)  + F_n(t_1) S_n^\prime(t_2) }\nonumber\\
&\equiv \frac{N(t_1,t_2)}{D(t_1,t_2)},\label{ACP1}
\end{align}
where $F_h(t_1)\equiv e^{-\Gamma_D t_1} \sinh{\frac{1}{2}\Delta \Gamma_D t_1}$, $F_n(t_1) \equiv e^{-\Gamma_D t_1} \sin{\Delta m_D t_1}$, $\Gamma \equiv \pqty{\Gamma_H + \Gamma_L}/2, \Delta \Gamma \equiv \Gamma_H - \Gamma_L$ and $\Delta m \equiv m_H - m_L.$ The term proportional to $\abs{g_{+,D}(t_1)}^2$ arises from the interference of amplitudes corresponding to all possible pairs of paths, both of which involve the direct decay of $D^0$ without oscillation, while the term proportional to $\abs{g_{-,D}(t_1)}$ is induced by the interference of amplitudes corresponding to all possible paths in which $D^0$ oscillates to $\bar{D}^0$ before decaying. The terms proportional to $F_h(t_1)$ and $F_n(t_1)$ represent the CP violation induced by the interference between one path in which $D^0$ oscillates to $\bar{D}^0$ and subsequently decays, and another path in which $D^0$ decays directly without oscillation.
The coefficients $C_+(t_2)$ and $C_-(t_2)$ are calculable and will be introduced in the next section.

\section{Numerical analysis}
\subsection{$D^0(t_1) \to \pi^0 K(t_2)  \to \pi^0 (\pi^-\ell^+\nu_\ell) $}

The first case we focus on is $D^0(t_1) \to \pi^0 K(t_2) \to \pi^0(\pi^-\ell^+\nu_\ell)$, with its decay process depicted in the left panel of Fig.~\ref{fig:decay}, where the spectator meson $\pi^0$ is omitted. The amplitudes in the first-level decay $D \to \pi^0 K$ are interconnected through several parameters
\begin{align}
&\frac{A(\bar{D}^0 \to \pi^0 K^0)}{A(D^0 \to \pi^0 \bar{K}^0)} = e^{i\omega_1}, \quad \omega_1 = \arg \frac{V_{cs}V^\star_{ud}}{V^\star_{cs}V_{ud}},\nonumber\\
&\frac{A(D^0 \to \pi^0 K^0)}{A(D^0 \to \pi^0\bar{K}^0)} = r e^{i(\delta_2 + \omega_2)}, \quad \omega_2 = \arg \frac{V^\star_{cd}V_{us}}{V^\star_{cs} V_{ud}},\label{amplitudes}
\end{align}
where $\omega_{1,2}$ are the weak phases, $r$ is the magnitude ratio, and $\delta_2$ is the strong phase. Under the U-spin symmetry hypothesis, we have approximately $A(D^0 \to \pi^0 K^0)/A(D^0 \to \pi^0 \bar{K}^0)$ = $-\tan^2{\theta_C}$~\cite{CLEO:2007rhw,Rosner:2006bw}, where $\theta_C$ is the Cabibbo angle, which is also accepted by the factorization-assisted topological amplitude approach~\cite{Li:2012cfa,Qin:2013tje,Qin:2021tve}. Therefore, we adopt $r = \abs{(V_{cd}^\star V_{us})/(V_{ud} V_{cs}^\star)}$ and $\delta_2 = 0$ for the numerical analysis, with the value of $r$ being provided in~{\bf{Table}}~\ref{input}. We neglect the direct CP violation in $K^0$ decays, i.e., $\braket{\pi^-\ell^+\nu_\ell}{K^0} = \braket{\pi^+\ell^-\bar{\nu}_\ell}{\bar{K}^0}$. The CP violation associated with coefficients $C_+(t_2)$ and $C_-(t_2)$ is induced by the interference between amplitudes corresponding to all possible direct decay of the $D^0$ meson, $D^0 \to K^0 (\bar{K}^0)$, and by the interference between amplitudes corresponding to all oscillation-mediated paths, $D^0 \to \bar{D}^0 \to K^0(\bar{K}^0)$, respectively. They are calculated to be
\begin{align}
&C_{+,1}(t_2) =  \ re^{-\Gamma_K t_2}\cos{\delta_2}\Big\{4{\rm{Re}}(\epsilon_K)\sinh{\frac{1}{2}\Delta \Gamma_K t_2} \cos{(\omega_2-\phi_K)}+2\sin{\Delta m_K t_2}\nonumber\\
&\qquad \qquad \times\sin{(\omega_2-\phi_K)}  \Big\}, \\
&C_{+,2}(t_2) =  8{\rm{Re}}
(\epsilon_K)\abs{g_{-,K}(t_2)}^2,\label{Cpluscase1}\\
&C_-(t_2)=\ re^{-\Gamma_K t_2}\cos{\delta_2}\Big\{ 2\sin{\Delta m_K t_2}\sin{(\omega_2-\phi_K)} + 4\Big({\rm{Re}}(\epsilon_K) - 2 {\rm{Re}}(\epsilon_D)\Big)\nonumber\\
&\qquad \qquad\times \sinh{\frac{1}{2}\Delta \Gamma_K t_2 }\cos{(\omega_2 - \phi_K)}\Big\} -8\abs{g_{+,K}(t_2)}^2{\rm{Re}}(\epsilon_D),\label{Cminuscase1}
\end{align}
where the sum of $C_{+,1}(t_2)$ and $C_{+,2}(t_2)$ equals to $C_+(t_2)$. Based on the input parameter values in Table~\ref{input}, we obtain $\sin{(\omega_2 - \phi_K)} \approx 3.7 \times 10^{-3}$, $\sin{(\omega_1 - \omega_2 -\phi_D)} \approx -6.4 \times 10^{-4}$, $\sin{(\omega_1 - \phi_D - \phi_K)} \approx -3.1 \times 10^{-3}$, $\sin{(\omega_1 - 2\omega_2 - \phi_D + \phi_K)} \approx 4.4 \times 10^{-3}$ and $r^2 \approx 2.8 \times 10^{-3}$, which are of order $\mathcal{O}(10^{-3})$ or smaller. When multiplied with ${\rm{Re}}(\epsilon_K) \sim 10^{-3}$ and ${\rm{Re}}(\epsilon_D) \sim 10^{-4}$, their overall contributions are negligible. The contributions from $C_+, C_-$, as well as the double-mixing CP violation terms $S_{h,\rm{dm}}$ and $S_{n,\rm{dm}}$ within $S_h$ and $S_n$ in Eq. \eqref{ACP1}, with coefficients of order $\mathcal{O}(10^{-5})$ or smaller, where the coefficients refer to the parts excluding time-evolution factors such as $\abs{g_{-,K}(t_2)}^2$, are neglected, as these contributions are at least one order smaller than the leading contributions in $C_+$, $C_-$, $S_{h,{\rm{dm}}}$ and $S_{n, {\rm{dm}}}$. Therefore, the CP violation induced by $D^0 \to \bar{D}^0 \to \bar{K}^0 \to K^0$, as well as the CP violation induced by the interference between $D^0 \to K^0$ and $D^0 \to \bar{D}^0 \to \bar{K}^0 \to K^0$ are neglected. For the non-double-mixing CP violation terms in $S_h$ and $S_n$, given their minimal contributions, no approximations are made to allow for a full analysis of their behavior. In Eq. \eqref{ACP1}, only the leading terms in the denominators are retained, with terms involving the aforementioned small quantities omitted.

The double-mixing CP violation terms $S_{h,\rm{dm}}$ and $S_{n,\rm{dm}}$ are calculated as
\begin{align}
& S_{n,\rm{dm}}(t_2) = \frac{e^{-\Gamma_K t_2}}{2}\Big\{4\Big( {\rm{Re}}{(\epsilon_K)} - {\rm{Re}}(\epsilon_D)\Big)\sin{\Delta m_K t_2}\cos{(\omega_1-\phi_D -\phi_K)} \nonumber\\
&\qquad \qquad \quad - 2\sinh{\frac{1}{2}\Delta \Gamma_K t_2}\sin{(\omega_1 - \phi_D - \phi_K)}\Big\},\\
&S_{h,\rm{dm}}(t_2) = \frac{e^{-\Gamma_K t_2}}{2}\Big\{4\Big( {\rm{Re}}{(\epsilon_K)} - {\rm{Re}}(\epsilon_D)\Big)\sinh{\frac{1}{2}\Delta\Gamma_K t_2}\cos{(\omega_1-\phi_D-\phi_K)}\nonumber\\
&\qquad \qquad \quad +2\sin{\Delta m_K t_2}\sin{(\omega_1-\phi_D-\phi_K)}\Big\}.
\end{align}
They are induced by the interference between $D^0 \to \bar{K}^0 \to K^0$ and $D^0 \to \bar{D}^0 \to K^0$, while contributions from the interference between $ D^0 \to K^0 $ and $D^0 \to \bar{D}^0 \to \bar{K}^0 \to K^0$ are neglected due to suppression by terms of $ r^2$. The remaining components of $S_n(t_2)$ and $S_h(t_2)$ are given by
\begin{align}
& S_{n,\rm{others}}(t_2) = -2re^{-\Gamma_K t_2}\cosh{\frac{1}{2}\Delta \Gamma_K t_2}\cos{\delta_2}\sin{\pqty{\omega_1-\omega_2-\phi_D}},\\
&S_{h,\rm{others}}(t_2) = r\cos{\delta_2}\cos{(\omega_1-\omega_2-\phi_D)}\Big[-4{\rm{Re}}(\epsilon_D)\abs{g_{+,K}(t_2)}^2\nonumber\\
& \qquad \qquad \quad +4\Big(2{\rm{Re}}(\epsilon_K)  - {\rm{Re}}(\epsilon_D)\Big)\abs{g_{-,K}(t_2)}^2 \Big],
\end{align}
which are induced by the interference between $D^0 \to \bar{D}^0 \to K^0$ and $D^0 \to K^0$, as well as the interference between $D^0 \to \bar{K}^0 \to K^0$ and $D^0 \to \bar{D}^0 \to \bar{K}^0 \to K^0$. The terms in the denominator of Eq. \eqref{ACP1} are given by
\begin{align}
&C^\prime_+(t_2) = \ 2\abs{g_{-,K}(t_2)}^2+ 2re^{-\Gamma_K t_2}\sinh{\frac{1}{2}\Delta \Gamma_K t_2}\cos{\delta_2} \cos{(\omega_2-\phi_K)},\\
&C^\prime_-(t_2) = 2\abs{g_{+,K}(t_2)}^2 +2re^{-\Gamma_K t_2}\sinh{\frac{1}{2}\Delta \Gamma_K t_2 }\cos{\delta_2}\cos{(\omega_2 - \phi_K)} ,\\
&S^\prime_h(t_2) = e^{-\Gamma_K t_2}\sinh{\frac{1}{2}\Delta\Gamma_K t_2}\cos{(\omega_1-\phi_D-\phi_K)}+2re^{-\Gamma_K t_2}\cosh{\frac{1}{2}\Delta \Gamma_K t_2}\cos{\delta_2}\nonumber\\
&\qquad \qquad \times\cos{(\omega_1-\omega_2-\phi_D)},\\
&S^\prime_n(t_2) = e^{-\Gamma_K t_2}\sin{\Delta m_K t_2}\cos{(\omega_1 - \phi_D -\phi_K)}.
\end{align}

It is important to note that, according to Eq. \eqref{ACP1}, the CP violation corresponding to each contribution is defined as
\begin{align}
A_{C_+}(t_1,t_2) &\equiv \frac{C_+(t_2)\abs{g_{+,D}(t_1)}^2}{D(t_1,t_2)},\label{C+}\\
A_{C_-}(t_1,t_2) &\equiv \frac{C_-(t_2)\abs{g_{-,D}(t_1)}^2}{D(t_1,t_2)},\label{C-}\\
A_{S_h}(t_1,t_2) & \equiv \frac{e^{-\Gamma_D t_1}\sinh{\frac{1}{2}\Delta \Gamma_D t_1}S_h(t_2)}{D(t_1,t_2)},\label{Sh}\\
A_{S_n}(t_1,t_2) & \equiv \frac{e^{-\Gamma_D t_1}\sin{\Delta m_D t_1}S_n(t_2)}{D(t_1,t_2)}.\label{Sn}
\end{align}

\begin{figure}[htbp]
    \centering
    \includegraphics[keepaspectratio,width=5.2cm]{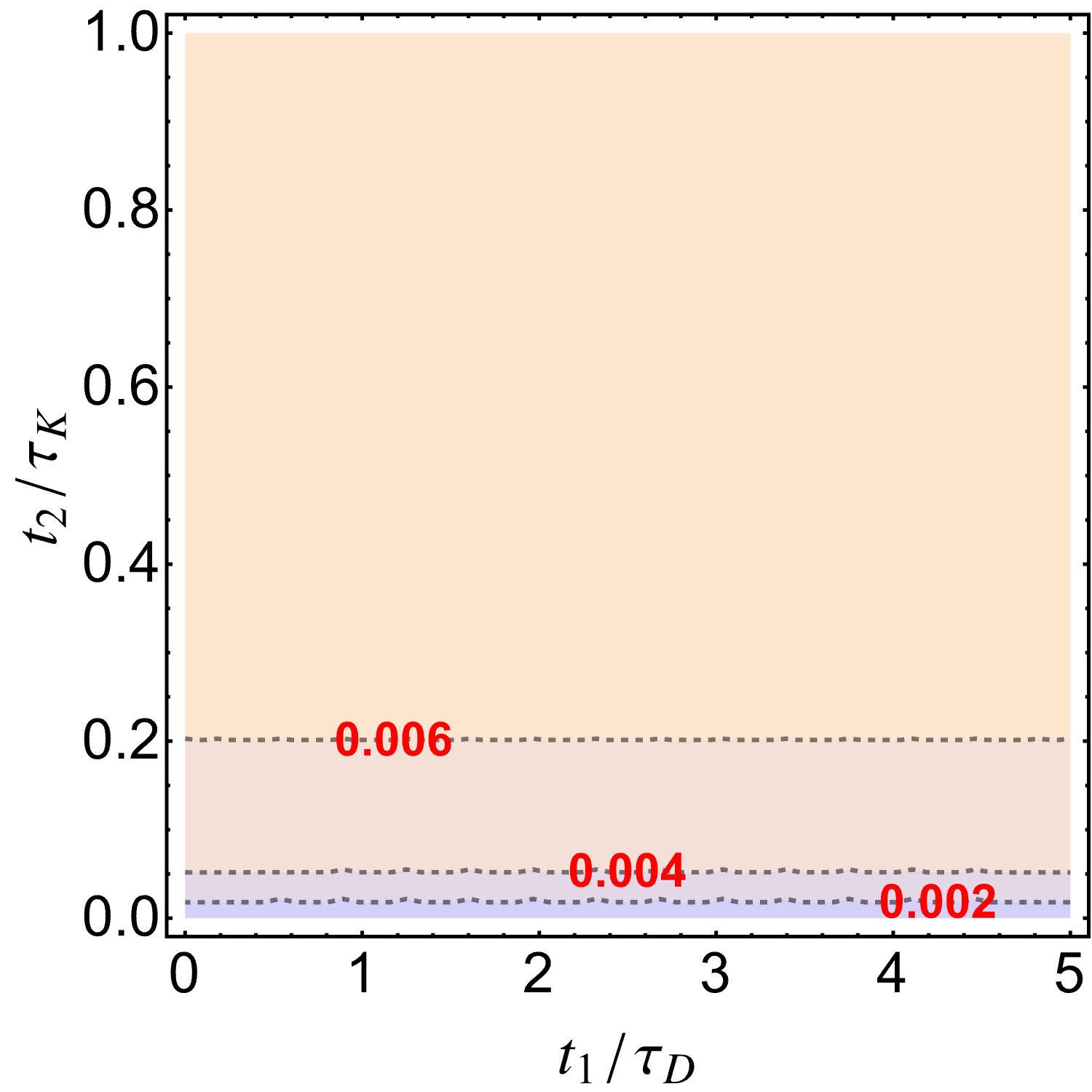}
    \hspace{0.1cm}
    \includegraphics[keepaspectratio,width=5.2cm]{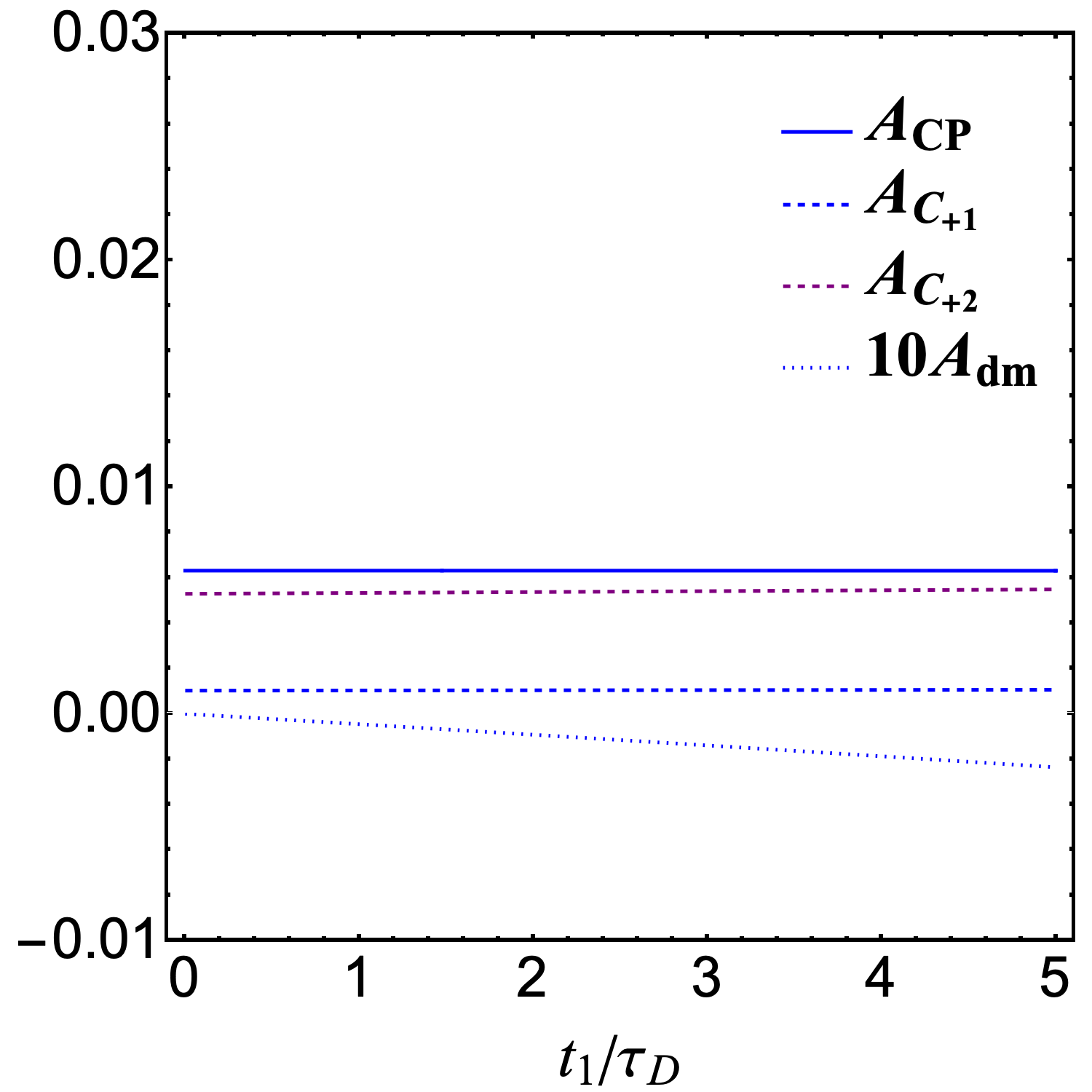}
     \hspace{0.1cm}
    \includegraphics[keepaspectratio,width=5.2cm]{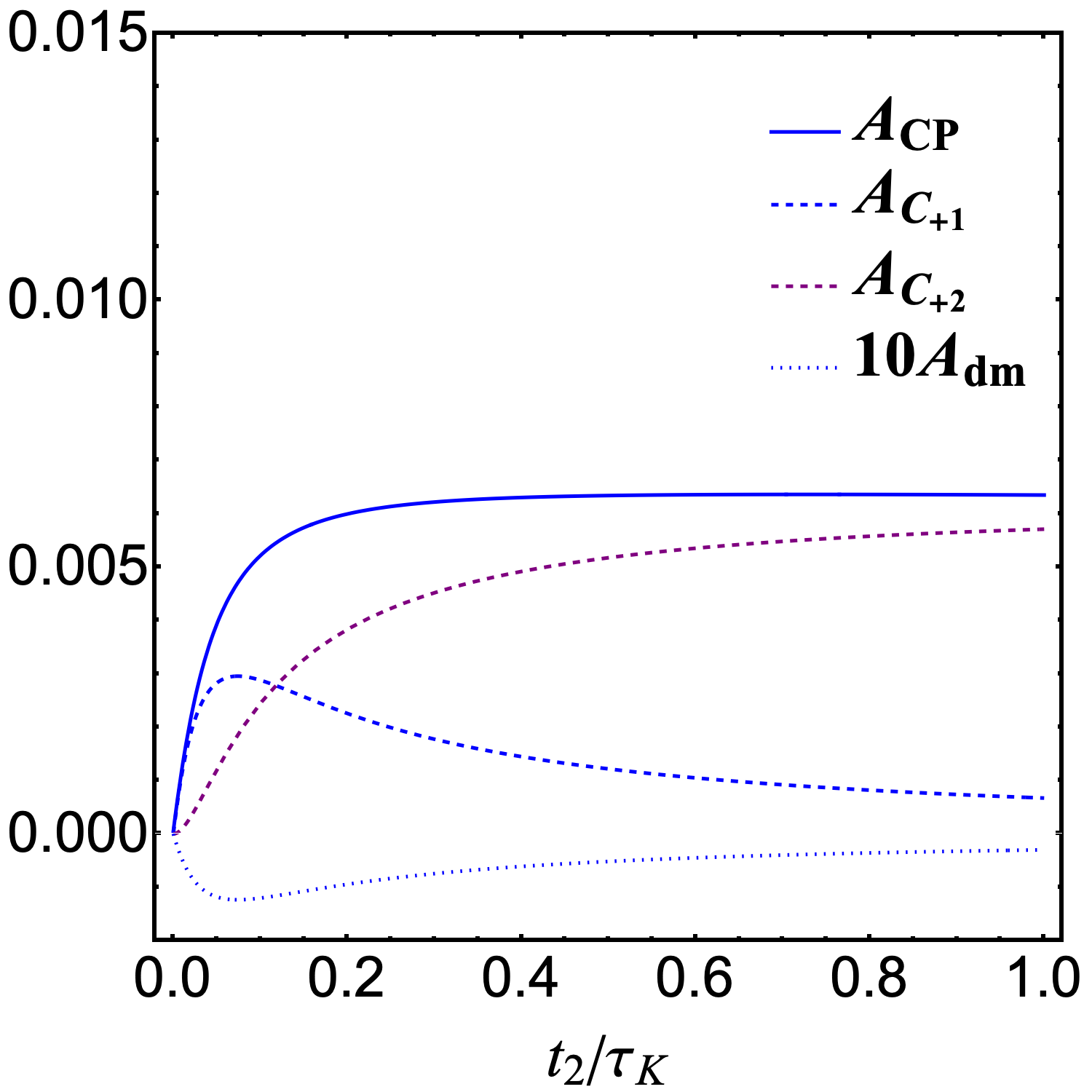}
    \caption{Time dependence of the CP asymmetry $A_{\rm CP}$ in $D(t_1) \to \pi^0 K(t_2) \to \pi^0 (\pi^-\ell^+\nu_\ell)$. The left panel displays the two-dimensional time-dependent total CP violation. The middle panel and the right panel display the dependence on $t_1$ (with $t_2$ integrated from 0 to $\tau_{K}$) and $t_2$ (with $t_1$ integrated from $0$ to $5\tau_D$), respectively.}\label{fig:D1}
\end{figure}

To provide a more intuitive analysis, we performed a numerical analysis of the contributions to the total CP violation in this decay channel using the values listed in {\bf{Table}}~\ref{input} as input parameters. The results are shown in Fig.~\ref{fig:D1}. The left panel illustrates the two-dimensional time-dependent total CP violation, with its peak value exceeding $5 \times 10^{-3}$, which occurs when $t_2$ is large. After integrating $ A_{CP}(t_1, t_2) $ in Eq. \eqref{ACP1} with respect to $ t_2 $ from 0 to $\tau_K$, where $\tau_K \equiv 1/\Gamma_K = 2/(\Gamma_{K_S} + \Gamma_{K_L}),$ the resulting CP violation as a function of $t_1$ is depicted in the middle panel. Among the contributions to the total CP violation, $A_{C_+}$, which is proportional to $\abs{g_{+,D}(t_1)}^2$, is the dominant term. The contribution to $A_{C_+}$ can be divided into two components. The first ($A_{C_{+,1}}$) refers to Eq.~\eqref{C+} where $C_+(t_2)$ is replaced by $C_{+,1}(t_2)$. This contribution arises from CP violation induced by the interference between $D^0 \to K^0$ and $D^0 \to \bar{K}^0 \to K^0$, which is associated with the interference between the Cabibbo-favored and doubly-Cabibbo-suppressed amplitudes involving $K^0 - \bar{K}^0$ mixing~\cite{Yu:2017oky}. The second one ($A_{C_{+,2}}$) refers to Eq.~\eqref{C+} where $C_+(t_2)$ is replaced by $C_{+,2}(t_2)$, and originates from CP violation directly induced by $D^0 \to \bar{K}^0 \to K^0$. The latter contribution is approximately five times larger than the former, with both of the order $\mathcal{O}(10^{-3})$. The dominant contribution, $A_{C_{+,2}}$, remains almost time-independent, as the denominator in Eq. \eqref{ACP1} is primarily proportional to $\abs{g_{+,D}(t_1)}^2$, resulting in the time-dependent factor associated with $t_1$ canceling out. Similarly, the sub-leading contribution, $A_{C_{+,1}}$, exhibits negligible time dependence for the same reason. The double-mixing CP violation $A_{h,\rm{dm}}$ increases linearly with time $t_1$, reaching $\mathcal{O}(10^{-4})$, which is one order of magnitude smaller than the sub-leading term. The double-mixing CP violation $A_{n,\rm{dm}}$ and other terms $A_{\rm{others}}$ are exceedingly small, on the order of $\mathcal{O}(10^{-6})$ and $\mathcal{O}(10^{-7})$, respectively. Integrating $t_1$ from 0 to $5\tau_D$, the right panel shows that when $t_2 < 0.1\tau_K$, the contributions from $A_{C_{+,1}}$ and $A_{C_{+,2}}$ are comparable ($\sim 10^{-3}$), with the former slightly larger than the latter. Subsequently, $A_{C_{+,2}}$ increases and then remains constant, while $A_{C_{+,1}}$ decreases and stabilizes. In the stable state, $A_{C_{+,2}}$ is approximately five times larger than $A_{C_{+,1}}$. The double-mixing CP violation $A_{h,\rm{dm}}$ is one order smaller than the two dominant contributions, but still two orders larger than $A_{n,\rm{dm}}$. This discrepancy arises because, although the magnitude of the coefficients in front of $\sin{x_K \Gamma_K t_2}$ and $\sinh{y_K \Gamma_K t_2}$ are nearly identical in both $A_{h,dm}$ and $A_{n,dm}$, the two terms have the same sign in $A_{h,dm}$, while they have opposite signs in $A_{n,dm}$. Both $A_{C_-} (\sim 10^{-7})$ and the other contributions, $A_{\rm{others}} (\sim 10^{-5})$, decrease over time.

\begin{table}[!h]
\begin{center}
\caption{The input parameters and their values, with $x_M \equiv \Delta m_M/\Gamma_M$ and $y_M \equiv \Delta\Gamma_M / (2\Gamma_M)$, respectively.}\label{input}
\begin{tabular}{||cr|cr||}
\hline
 \textbf{Parameter}    &  \textbf{Value} & \textbf{Parameter}    &  \textbf{Value} \\
\hline\hline
  $|q_K/p_K|$   &    $0.996774\pm 0.000016$~\cite{ParticleDataGroup:2024cfk} & $x_K$ & $0.946\pm 0.002$~\cite{ParticleDataGroup:2024cfk} \\
  $|q_D/p_D|$   &    $0.9997$~\cite{Li:2022jxc}  & $x_D$ & $\pqty{0.21^{+0.04}_{-0.07}}\%$~\cite{Li:2022jxc} \\
  $\phi_K$      &    $\pqty{0.176\pm 0.001}^\circ$~\cite{ParticleDataGroup:2024cfk} & $y_K$ & $-0.9965\pm 0.0006$~\cite{ParticleDataGroup:2024cfk} \\
  $\phi_D$      &    $-\pqty{3.1^{+0.3}_{-0.4}}^\circ \times 10^{-3} $~\cite{Li:2022jxc} & $y_D$ & $\pqty{0.52\pm0.03}\%$~\cite{Li:2022jxc} \\
  $\omega_1$ & $(-0.00377^{+0.000124}_{-0.000116})^\circ $~\cite{ParticleDataGroup:2024cfk}& $\omega_2$ & $179.96^\circ$~\cite{ParticleDataGroup:2024cfk}\\
  $\abs{(V_{cd}^\star V_{us})/(V_{ud}V_{cs}^\star)}$ & 0.0534 ~\cite{CLEO:2007rhw}& &\\
\hline
\end{tabular}
\end{center}
\end{table}

We define the absolute branching ratio for $D^0(t_1) \to \pi^0 K(t_2) \to \pi^0 (\pi^- \ell^+ \nu_\ell)$ as 
\begin{align}
\frac{\int_0^{t_1}\int_0^{t_2} \Gamma[D^0(t_1) \to \pi^0 K(t_2) \to \pi^0 (\pi^- \ell^+ \nu_\ell)]}{\int_0^{t_1}\int_0^{t_2} \Gamma[D^0(t_1) \to \pi^0 K(t_2) \to \pi^0 (\pi^+ \pi^-)]}\times \mathcal{B}(D^0 \to \pi^0 K^0_S)\times \mathcal{B}(K^0_S \to \pi^+ \pi^-)\times A(t_2),\label{BrD1}
\end{align}
where the branching ratios $\mathcal{B}(D^0 \to \pi^0 K^0_S)$ and $\mathcal{B}(K^0_S \to \pi^+ \pi^-)$ are 1.24\% and 69.2\%~\cite{ParticleDataGroup:2024cfk}, respectively. $A(t_2)$ represents the fraction of $K^0_S$ decays that have occurred by time $t_2$, as defined in Eq.~\eqref{At}. The results of the integration over the two time parameters, $t_1$ and $t_2$, are shown in~{\bf{Table}}~\ref{BRD}. The absolute branching ratio is sensitive to the integration range of the time parameter $t_2$ for $K$ meson, but nearly insensitive to the integration range of the time parameter $t_1$ for $D$ meson. The absolute branching ratio increases as the integration range of $t_2$ grows.

\begin{table}[h!]
\setlength{\tabcolsep}{20pt}
\centering
\caption{The absolute branching ratio for $D^0(t_1) \to \pi^0 K(t_2) \to \pi^0 (\pi^- \ell^+ \nu_\ell)$}\label{BRD}
\begin{tabular}{ccc}
\hline
$t_1/\tau_{D}$ & $t_2/ \tau_K$ & $\rm{Br}$\\
\hline
5 & 1.5 & $1.69\times 10^{-5}$\\
5 & 1 & $7.55\times 10^{-6}$\\
5 & 0.5 & $1.63\times 10^{-6}$\\
\hline
\end{tabular}
\end{table}

\subsection{$D^0(t_1) \to \pi^0 K(t_2)  \to \pi^0 (\pi^+\ell^-\bar{\nu}_\ell) $}

In contrast to the previous decay channel, the semi-leptonic final state of the neutral kaon is changed from $\pi^-\ell^+\nu_\ell$ to $\pi^+\ell^-\bar{\nu}_\ell$. The decay process for this channel is illustrated in the right panel of Fig. \ref{fig:decay}. By substituting Eq. \eqref{amplitudes} into Eq. \eqref{ACP1}, the time-dependent contributions to the overall CP violation as a function of $t_2$ are given by
\begin{eqnarray}
C_+(t_2) &=& re^{-\Gamma_K t_2}\cos{\delta_2} \Big\{-4{\rm{Re}}(\epsilon_K)\sinh{\frac{1}{2}\Delta \Gamma_K t_2} \cos{(\omega_2 -\phi_K)} - 2\sin{\Delta m_K t_2}\nonumber\\
&&\times \sin{(\omega_2 -\phi_K)} \Big\},\\
C_-(t_2) &=& -2re^{-\Gamma_K t_2}\sin{\Delta m_K t_2 }\cos{\delta_2} \sin{(\omega_2 - \phi_K)} -8\Big( {\rm{Re}}(\epsilon_K) + {\rm{Re}}(\epsilon_D)\Big)\abs{g_{-,K}(t_2)}^2  \nonumber\\
&& -4re^{-\Gamma_K t_2}\cos{\delta_2}\Big( {\rm{Re}}(\epsilon_K) + 2{\rm{Re}}(\epsilon_D)\Big)\sinh{\frac{1}{2}\Delta \Gamma_K t_2}\cos{(\omega_2-\phi_K)},\\
S_{n,\rm{dm}}(t_2) &=& \frac{e^{-\Gamma_K t_2}}{2}\Big\{-2\sinh{\frac{1}{2}\Delta\Gamma_K t_2}\sin{(\omega_1-\phi_D-\phi_K)} + 4 \Big( {\rm{Re}}(\epsilon_K) + {\rm{Re}}(\epsilon_D) \Big)\nonumber\\
&&\times \sin{\Delta m_K t_2}\cos{(\omega_1-\phi_D-\phi_K)} \Big\},\\
S_{n,\rm{others}}(t_2) &=& -2re^{-\Gamma_K t_2}\cosh{\frac{1}{2}\Delta \Gamma_K t_2}\cos{\delta_2}\sin{(\omega_1-\omega_2-\phi_D)},\\
S_{h,\rm{dm}}(t_2) &=& \frac{e^{-\Gamma_K t_2}}{2} \Big\{-4 \Big( {\rm{Re}}(\epsilon_K) + {\rm{Re}}(\epsilon_D)\Big)\sinh{\frac{1}{2}\Delta\Gamma_K t_2}\cos{(\omega_1-\phi_D-\phi_K)}\nonumber\\
&& -2\sin{\Delta m_K t_2}\sin{(\omega_1-\phi_D-\phi_K)} \Big\},\\
S_{h,\rm{others}}(t_2) &=& r\cos{\delta_2}\cos{(\omega_1-\omega_2-\phi_D)}\Big\{-4{\rm{Re}}(\epsilon_D)\abs{g_{+,K}(t_2)}^2\nonumber\\
&& -4\Big(  2{\rm{Re}}(\epsilon_K) - {\rm{Re}}(\epsilon_D)\Big)\abs{g_{-,K}(t_2)}^2 \Big\}.
\end{eqnarray}

Similar to the previous decay channel, terms with coefficients of $\mathcal{O}(10^{-5})$ or smaller are neglected, and thus the CP violation induced by $D^0 \to K^0 \to \bar{K}^0$ and $D^0 \to \bar{D}^0 \to \bar{K}^0$, as well as the CP violation incuded by the interference between $D^0 \to \bar{D}^0 \to \bar{K}^0$ and $D^0 \to K^0 \to \bar{K}^0$ are omitted. After disregarding negligible terms, $C_+(t_2)$ originates from the interference between $D^0 \to K^0 \to \bar{K}^0$ and $D^0 \to \bar{K}^0$, which is related to the effect discussed in~\cite{Yu:2017oky}, while $C_-(t_2)$ arises from two distinct contributions. The first contribution is associated with interference between $D^0 \to \bar{D}^0 \to \bar{K}^0$ and $D^0 \to \bar{D}^0 \to K^0 \to \bar{K}^0$. The second contribution is induced by the path $D^0 \to \bar{D}^0 \to K^0 \to \bar{K}^0$. The $t_2$-dependent factors $S_{n,\rm{dm}}(t_2)$ and $S_{h,\rm{dm}}(t_2)$, which are associated with the double-mixing CP violation, are primarily induced by the interference between the paths $D^0 \to \bar{D}^0 \to K^0 \to \bar{K}^0$ and $D^0 \to \bar{K}^0$. In these terms, the double-mixing of $D^0$ and $K^0$ manifests in the expression through the combination of the mixing angles $\phi_D + \phi_K$. The terms in the denominator of Eq. \eqref{ACP1} are given by
\begin{align}
&C^\prime_+(t_2) = 2\abs{g_{+,K}(t_2)}^2,\\
&C^\prime_-(t_2) = 2\abs{g_{-,K}(t_2)}^2 + 2re^{-\Gamma_K t_2}\sinh{\frac{1}{2}\Delta \Gamma_K t_2}\cos{\delta_2} \cos{(\omega_2 - \phi_K)},\\
&S^\prime_h(t_2) = e^{-\Gamma_K t_2} \sinh{\frac{1}{2}\Delta \Gamma_K t_2} \cos{(\omega_1 - \phi_D -\phi_K)} + 2r\abs{g_{+,K}(t_2)}^2 \cos{\delta_2} \cos{(\omega_1 - \omega_2 - \phi_D)},\\
&S^\prime_n(t_2) = -e^{-\Gamma_K t_2}\sin{\Delta m_K t_2} \cos{(\omega_1 - \phi_D - \phi_K)}.
\end{align}

\begin{figure}[htbp]
    \centering
    \includegraphics[keepaspectratio,width=5.2cm]{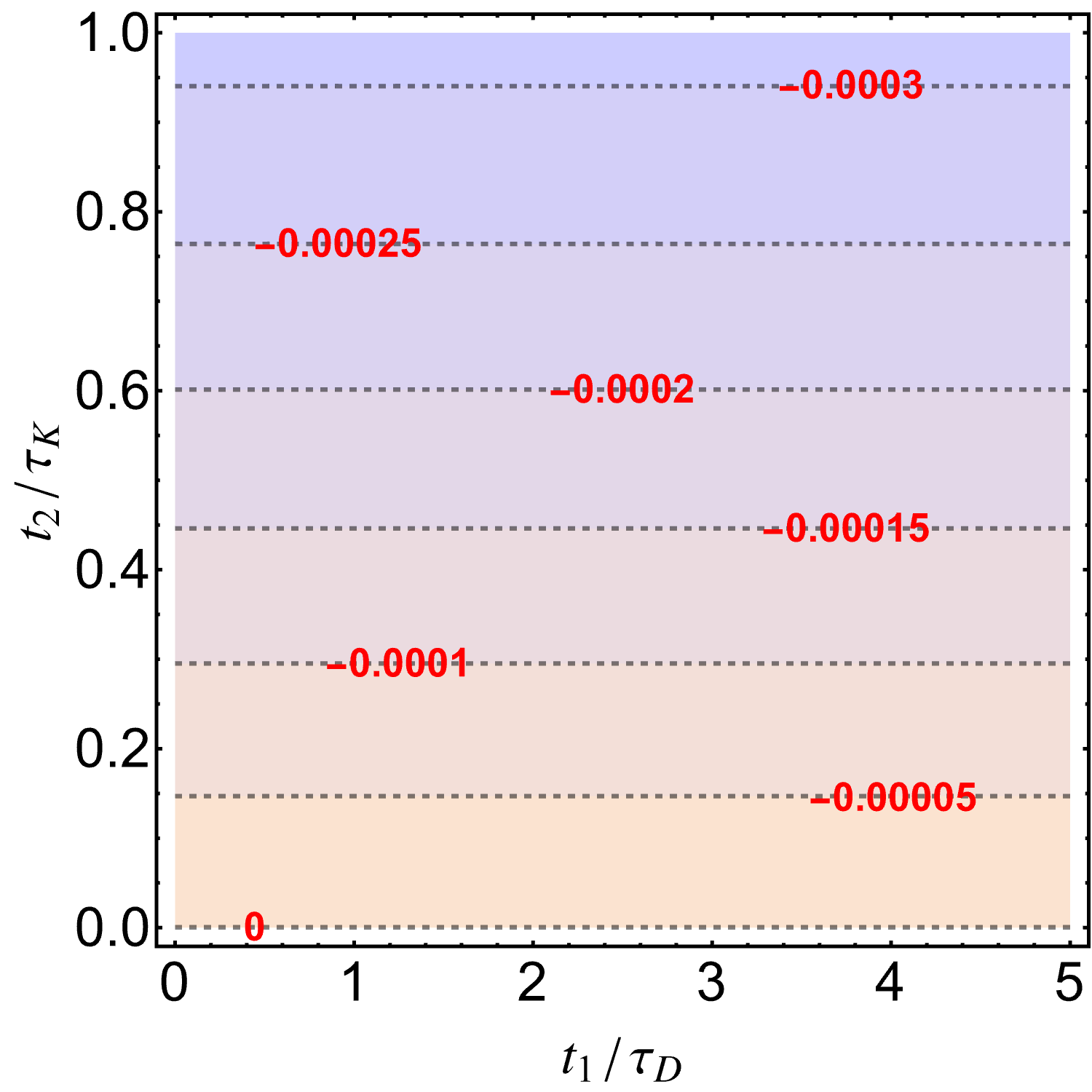}
    \hspace{0.1cm}
    \includegraphics[keepaspectratio,width=5.2cm]{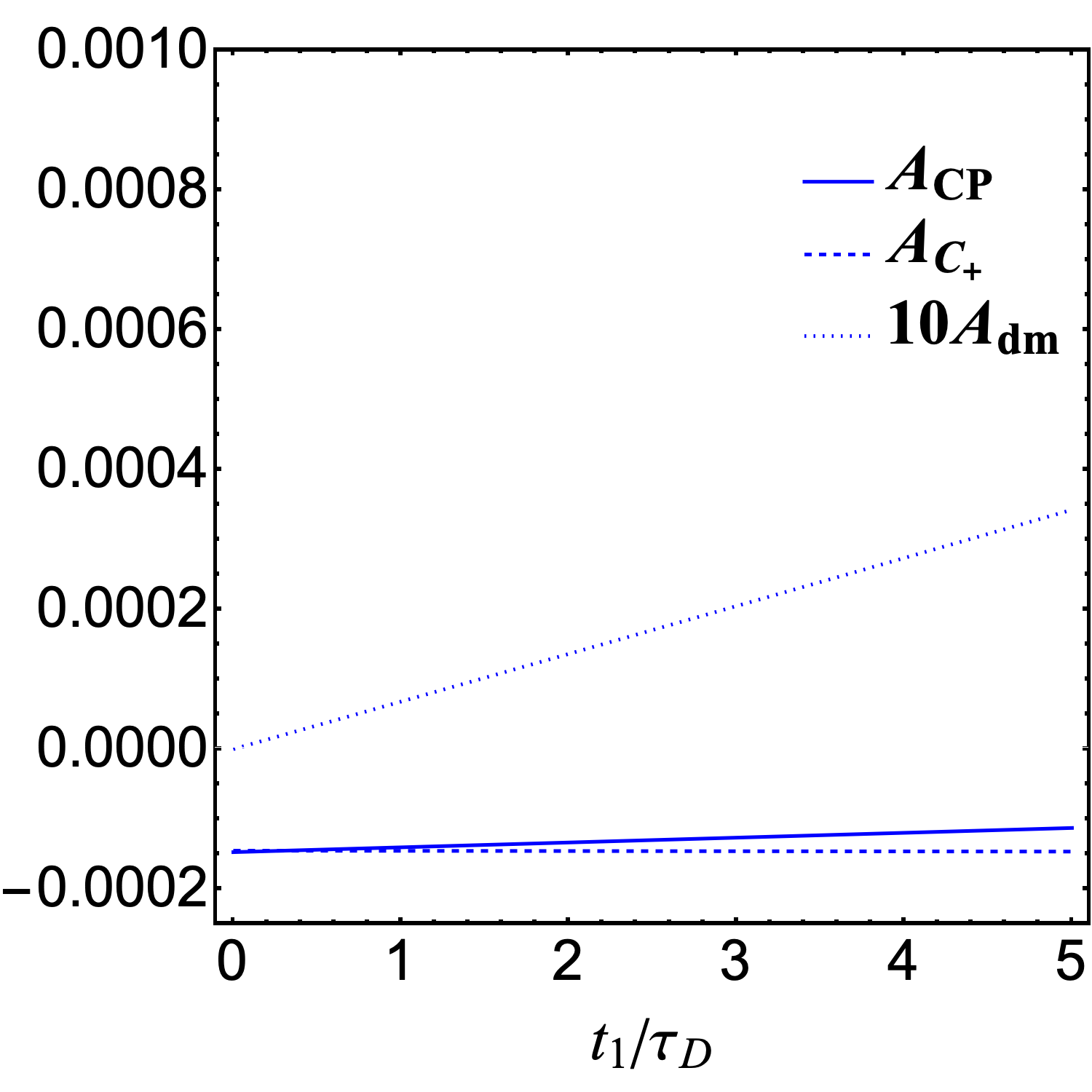}
     \hspace{0.1cm}
    \includegraphics[keepaspectratio,width=5.2cm]{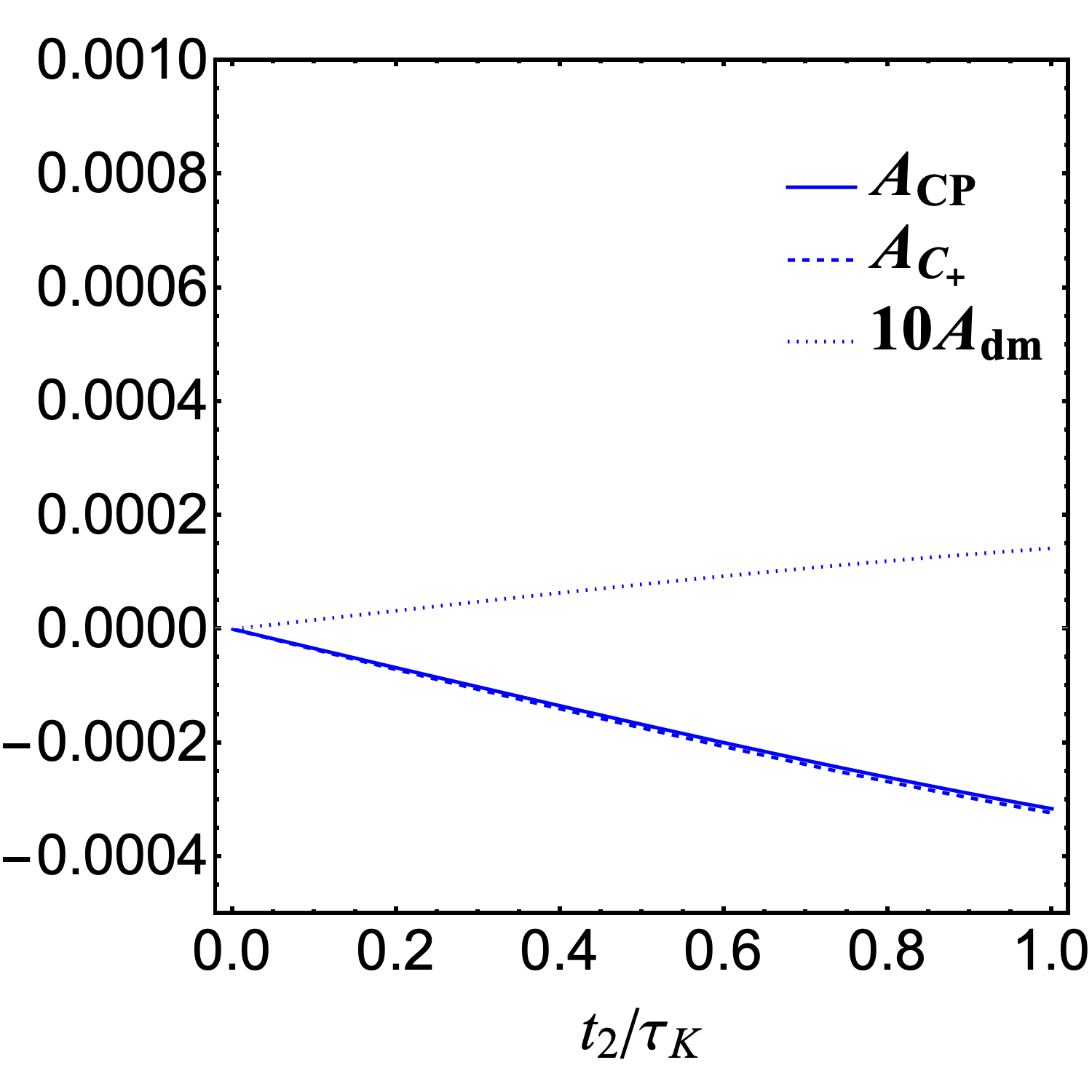}
    \caption{Time dependence of the CP asymmetry $A_{\rm CP}$ in $D(t_1) \to \pi^0 K(t_2) \to \pi^0 (\pi^+\ell^-\bar{\nu}_\ell)$. The left panel displays the two-dimensional time-dependent total CP violation. The middle panel and the right panel display the dependence on $t_1$ (with $t_2$ integrated from 0 to $\tau_{K}$) and $t_2$ (with $t_1$ integrated from $0$ to $5\tau_D$), respectively.}\label{fig:D1n}
\end{figure}

Analogous to the previous decay process, we conducted a two-dimensional time-dependent analysis of the total CP violation for this process, along with one-dimensional time-dependent analyses of individual CP violation contributions as functions of $t_1$ or $t_2$, obtained by integrating over the other time dimension ($t_2$ or $t_1$, respectively). The results are presented in Fig.~\ref{fig:D1n}. It is evident that the CP violation associated with this decay channel is generally one order of magnitude smaller than that in the preceding process, where the neutral $K$ meson decays to $\pi^-\ell^+\nu_\ell$. This discrepancy primarily arises because the denominator $D$ in Eq. \eqref{ACP1} for this decay process is approximately an order of magnitude larger than that for the process with the final state $\pi^-\ell^+\nu_\ell$. The two-dimensional time-dependent total CP violation (left panel) is positive, in contrast to the previous process, and increases with $t_2$. By integrating $t_2$ from 0 to $\tau_K$, the contributions to CP violation as a function of $t_1$ are shown in the middle panel. The dominant and sub-leading contributions are $A_{C_+}$ and $A_{S_h}$, with magnitudes of $\mathcal{O}(10^{-4})$ and $\mathcal{O}(10^{-5})$, respectively. The dominance of $A_{C_+}$ is primarily due to being the CP violation induced by the amplitudes corresponding to the $D^0$ meson that does not undergo oscillations. Such oscillations suppress CP violation by introducing time evolution factors like $\sinh{y_D\Gamma_D t_1}$ and $\sin{x_D \Gamma_D t_1}$, where the mixing parameters $x_D$ and $y_D$ for the neutral $D$ meson are very small, approximately $10^{-3}$. These time evolution factors significantly reduce the magnitudes of $A_{S_h}$ and $A_{S_n}$. Additionally, the CP violation $A_{C_-}$ induced by paths where both $D^0$ mesons undergo oscillations is greatly suppressed by factors such as $\abs{g_{-,D}(t_1)}^2 \propto (\cosh{y_D \Gamma_D t_1} - \cos{x_D \Gamma_D t_1})$, which further reduces CP violation. The $t_2$-dependent CP violation (with $t_1$ integrated from 0 to $5\tau_D$) is shown in the right panel. Both the dominant term $A_{C_+}$ and the sub-leading contribution $A_{S_h}$ increase with time, with the former exhibiting a greater increase. This is primarily due to the factor obtained after integrating the $t_1$-dependent part for $A_{C_+}$ being two to three orders of magnitude larger than that for $A_{S_h}$, which amplifies the overall time variation of $A_{C_+}$.

The absolute branching ratio of the decay channel $D^0(t_1) \to \pi^0 K(t_2) \to \pi^0 (\pi^+ \ell^- \bar{\nu}_\ell)$, defined similarly to Eq.~\eqref{BrD1}, is shown in~{\bf{Table}}~\ref{BRD2}. The absolute branching ratio follows a similar trend to the previous decay channel, being sensitive to variations in the integration range of the time parameter $t_2$ for the $K$ meson, and increasing as the range of $t_2$ grows. Overall, the absolute branching ratio is larger than that of the previous decay channel.

\begin{table}[h!]
\setlength{\tabcolsep}{20pt}
\centering
\caption{The absolute branching ratio for $D^0(t_1) \to \pi^0 K(t_2) \to \pi^0 (\pi^+ \ell^- \bar{\nu}_\ell)$}\label{BRD2}
\begin{tabular}{ccc}
\hline
$t_1/\tau_{D}$ & $t_2/ \tau_K$ & $\rm{Br}$\\
\hline
5 & 1.5 & $6.6\times 10^{-5}$\\
5 & 1 & $5.3\times 10^{-5}$\\
5 & 0.5 & $3.3 \times 10^{-5}$\\
\hline
\end{tabular}
\end{table}

Considering the CP violation defined by the differences of the summed branching ratios
\begin{align}
\frac{\mathcal{B}[D^0/\bar{D}^0 \to \pi^0 K \to \pi^0(\pi^-\ell^+\nu_\ell)] - \mathcal{B}[D^0/\bar{D}^0 \to \pi^0 K \to \pi^0(\pi^+\ell^-\bar{\nu}_\ell)]}{\mathcal{B}[D^0/\bar{D}^0 \to \pi^0 K \to \pi^0(\pi^-\ell^+\nu_\ell)] + \mathcal{B}[D^0/\bar{D}^0 \to \pi^0 K \to \pi^0(\pi^+\ell^-\bar{\nu}_\ell)]},\label{combine}
\end{align}
it allows the experiment to proceed without the need for flavor tagging of the initial neutral $D$ meson. The components of the total CP violation are calculated as
\begin{align}
&C_+(t_2)  \approx 8{\rm{Re}}(\epsilon_K)\abs{g_{-,K}(t_2)}^2 + 4re^{-\Gamma_K t_2}\sin{\Delta m_K t_2}\cos{\delta_2}\sin{(\omega_2-\phi_K)}\nonumber\\
&\qquad \qquad\ + 8re^{-\Gamma_K t_2}{\rm{Re}}(\epsilon_K)\sinh{\frac{1}{2}\Delta \Gamma_K t_2}\cos{\delta_2}\cos{(\omega_2-\phi_K)},\\
&C_-(t_2) \approx -8{\rm{Re}}(\epsilon_D)\abs{g_{+,K}(t_2)}^2 +4\abs{g_{-,K}(t_2)}^2\Big( 2{\rm{Re}}(\epsilon_K) + {\rm{Re}}(\epsilon_D)\Big)\nonumber\\
&\qquad \qquad+ re^{-\Gamma_K t_2}\Big\{8{\rm{Re}}(\epsilon_K)\sinh{\frac{1}{2}\Delta \Gamma_K t_2}\cos{\delta_2}\cos{(\omega_2-\phi_K)} \nonumber\\
&\qquad \qquad +4\sin{\Delta m_K t_2}\cos{\delta_2}\sin{(\omega_2-\phi_K)}\Big\},\\
&S_h(t_2) = e^{-\Gamma_K t_2}\Big\{4{\rm{Re}}(\epsilon_K)\sinh{\frac{1}{2}\Delta \Gamma_K t_2}\cos{(\omega_1-\phi_D-\phi_K)}+2\sin{\Delta m_K t_2}\times\nonumber\\
&\qquad \qquad  \sin{(\omega_1-\phi_D-\phi_K)} \Big\} +16r{\rm{Re}}(\epsilon_K)\abs{g_{-,K}(t_2)}^2\cos{\delta_2}\cos{(\omega_1-\omega_2-\phi_D)},\\
&S_n(t_2) = -4e^{-\Gamma_K t_2}{\rm{Re}}(\epsilon_D)\sin{\Delta m_K t_2}\cos{(\omega_1 - \phi_D -\phi_K)}.
\end{align}

The denominators in Eq. \eqref{ACP} corresponding to the first and second decay modes are denoted as $D_1$ and $D_2$, respectively. Thus, the denominator in Eq. \eqref{combine} corresponds to $D_1 + D_2$.

\begin{figure}[htbp]
    \centering
    \includegraphics[keepaspectratio,width=5.2cm]{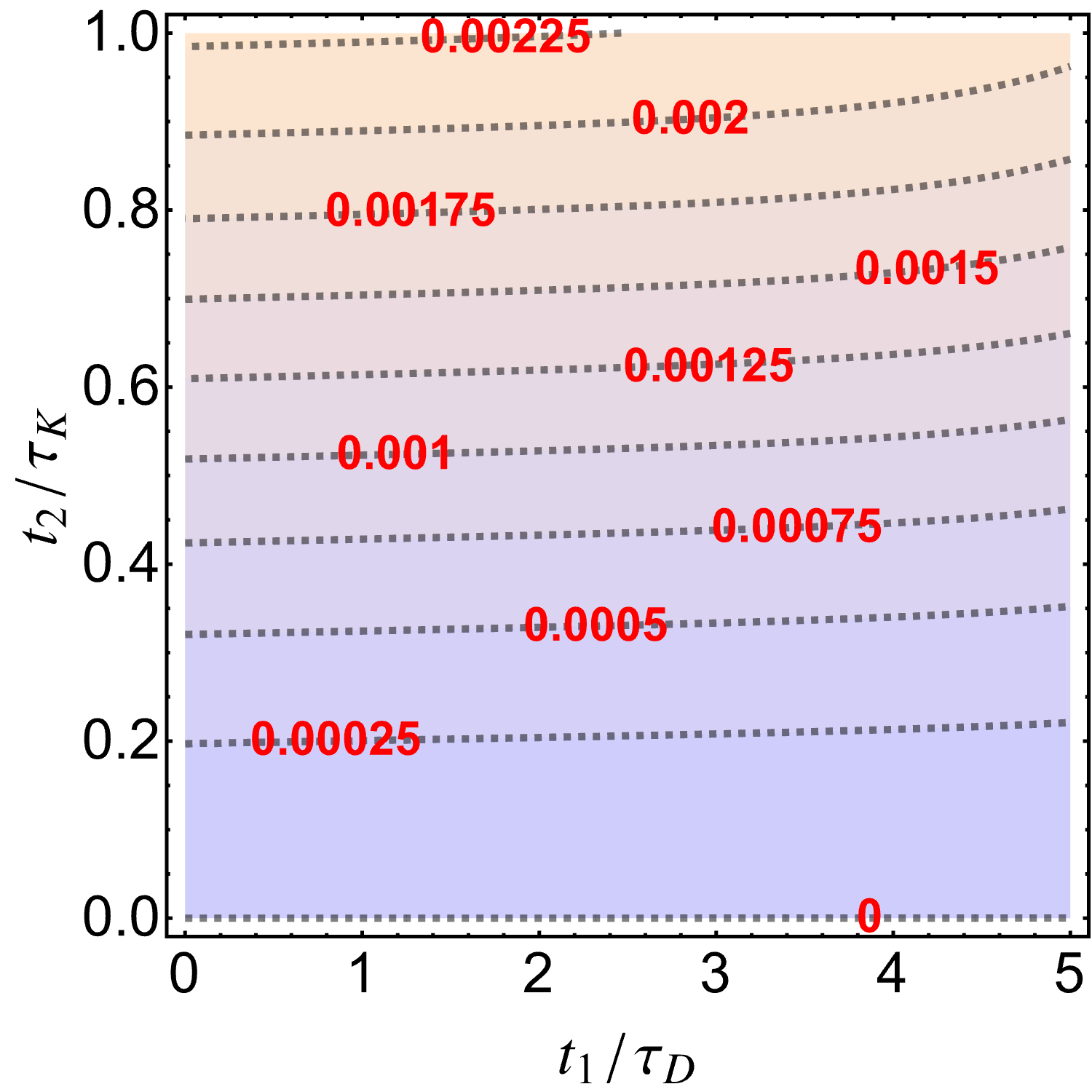}
    \hspace{0.1cm}
    \includegraphics[keepaspectratio,width=5.2cm]{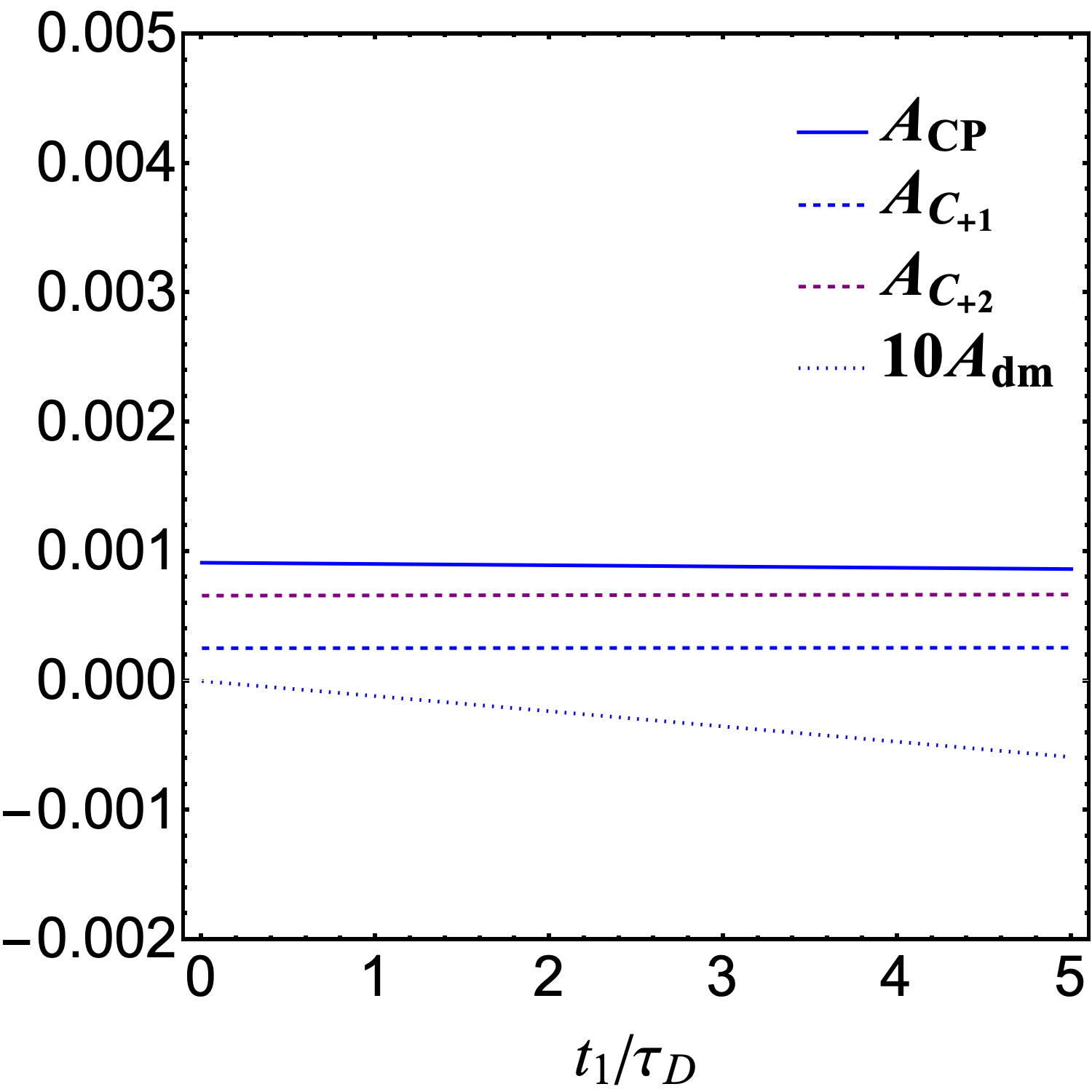}
     \hspace{0.1cm}
    \includegraphics[keepaspectratio,width=5.2cm]{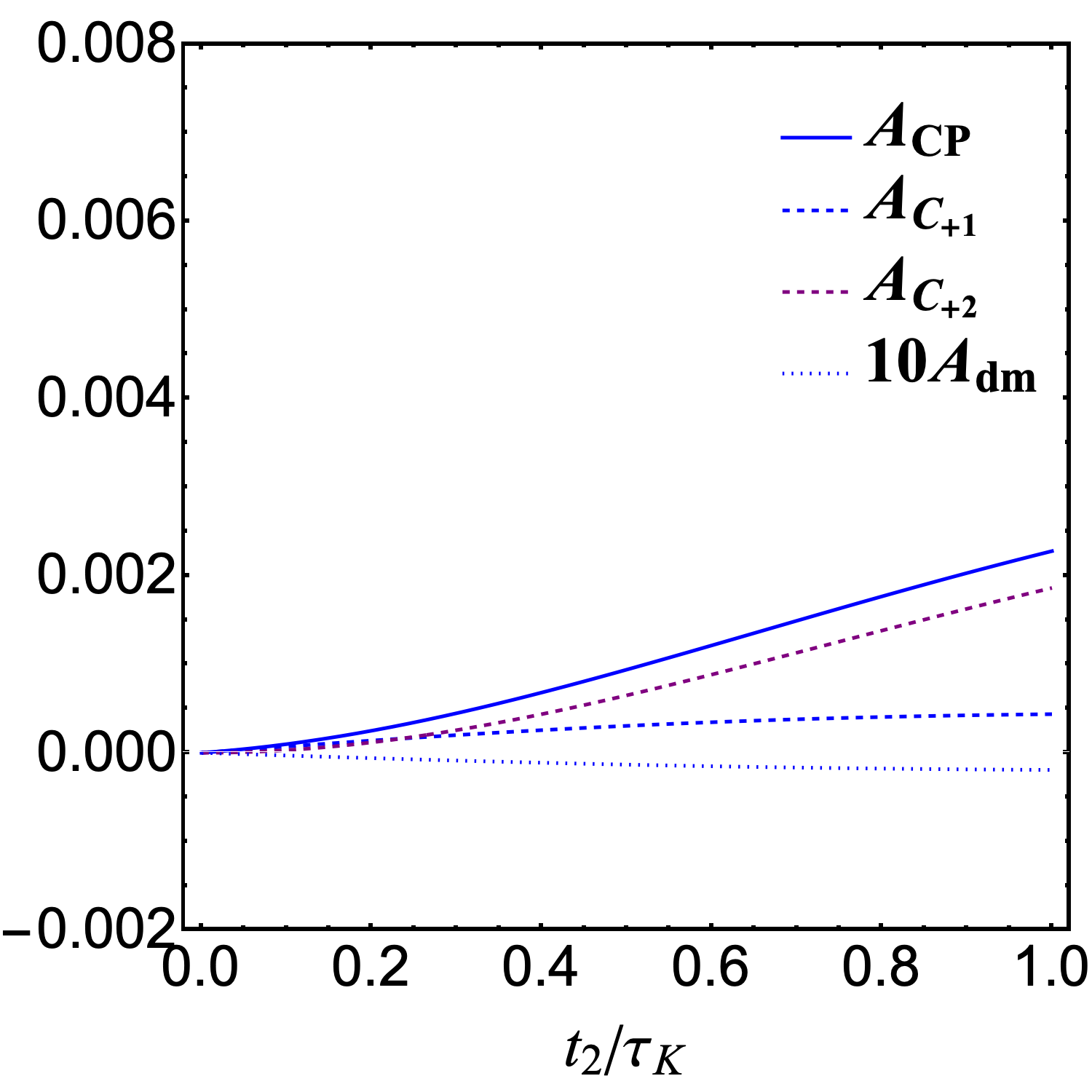}
    \caption{Time dependence of the CP asymmetry $A_{\rm CP}$ in $D^0(t_1) \to \pi^0 K(t_2) \to \pi^0(\pi\ell\nu)$ defined by Eq. \eqref{combine}. The left panel displays the two-dimensional time dependence. The middle panel and the right panel display the dependence on $t_1$ (with $t_2$ integrated from 0 to $\tau_K$) and $t_2$ (with $t_1$ integrated from 0 to $5\tau_D$), respectively.}\label{fig:DKminus(r)}
\end{figure}

The two-dimensional time-dependent total CP violation is shown in the left panel of Fig. \ref{fig:DKminus(r)}, with a peak being $10^{-3}$. Its magnitude lies between the results corresponding to the two decay channels, and it occurs when $t_2$ is large. By integrating $t_2$ from 0 to $\tau_K$, the contribution of each component to the total CP violation, as a function of $t_1$, is illustrated in the middle panel. The behavior of the individual components closely resembles that of the process $D^0(t_1) \to \pi^0 K(t_2) \to \pi^0(\pi^-\ell^+\nu_\ell)$, though their magnitudes are reduced by an order of magnitude. This reduction is primarily attributed to the increased denominator in Eq. \eqref{ACP}, which corresponds to the sum of the denominators for the two decay modes. The right panel shows the variation of CP violation with respect to $t_2$ after integrating $t_1$ from 0 to $5\tau_D$. The CP violation $A_{C_+}$, arising from the interference among all non-oscillating paths of the $D^0$ meson, remains the dominant contribution. It exhibits a significant increase, reaching $2 \times 10^{-3}$ at $t_2 = \tau_K$. In contrast, the sub-leading contribution $A_{S_h}$, is of the order $\mathcal{O}(10^{-5})$, two orders of magnitude smaller. Other contributions are heavily suppressed and can be safely neglected.

As the semileptonic $K^0$ decays have never been observed in heavy flavor experiments. The first step of measuring the discussed channels is to search $K^0\to \pi^-\ell^+\nu_\ell$ in experiments like BESIII, Belle (II) and LHCb. We attach relevant issues in Appendix~\ref{app}.

\section{Conclusion}
By this work, we have investigated the CP violation in the cascade decay $D^0(t_1) \to \pi^0 (K(t_2) \to \pi \ell \nu_\ell)$. Our analysis reveals that the CP violation arising from the interference of amplitudes corresponding to unmixed $D$ meson decays consistently dominates, irrespective of whether the contributions are $t_1$-dependent or $t_2$-dependent. The peak value of this dominant contribution reaches $5 \times 10^{-3}$, exceeding the sub-leading contribution, namely the double-mixing CP violation, by at least an order of magnitude. The choice of semileptonic final states proves to be critical: the CP violation contributions for the final state $\pi^- \ell^+ \nu_\ell$ are one to two orders of magnitude larger than those for $\pi^+ \ell^- \bar{\nu}_\ell $. By combining these two decay channels to measure the CP violation, experimental analyses can circumvent the need for flavor tagging of the initial $D$ meson, thereby avoiding the associated efficiency losses. The combined result represents an intermediate value between the CP violation observed in the two individual decay channels and is approximately half the value obtained for the channel with the $\pi^- \ell^+ \nu_\ell$ final state.

\section*{Acknowledgement.} 
The authors are grateful to Hai-Long Ma for useful discussions, and to Yin-Fa Shen for contribution in the early stage of this work.
This work is supported by Natural Science Foundation of China under grant No.~12375086.

\begin{appendix}
\section{Semileptonic $K^0$ decay} \label{app}
For the decay of a neutral $K$ meson into the semileptonic final state $\pi^- \ell^+ \nu_\ell$, the amplitude of the process is calculated as
\begin{align}
\mathcal{M}(K^0(t) \to \pi^- \ell^+ \nu_\ell) = g_{+,K}(t)\mathcal{A}_1.
\end{align}
Here, $\mathcal{A}_1$ denotes the amplitude for the decay $K^0 \to \pi^- \ell^+ \nu_\ell$. The corresponding decay width can be divided into three contributions:
\begin{align}
\Gamma(K^0(t) \to \pi^-\ell^+ \nu_\ell) = \Big(\frac{e^{-\Gamma_{K_S}t}}{4} + \frac{e^{-\Gamma_{K_L}t}}{4} + \frac{e^{-\Gamma_K t}}{2} \cos{\Delta m_K t}\Big) \abs{\mathcal{A}_1}^2,
\end{align}
where the first and second terms correspond to the contributions from the direct decays of $K^0_S$ and $K^0_L$, respectively, while the last term represents the $K^0_S - K^0_L$ interference. Consider the reference decay width
\begin{align}
\Gamma[K^0(t) \to \pi^+\pi^-] =& \abs{\mathcal{A}_2}^2 e^{-\Gamma_K t}\Big\{\cosh{\frac{1}{2}\Delta \Gamma_K t} + 2\Re(\epsilon_K)\cos{\Delta m_K t}\nonumber\\
& +\cos{\phi_K}\sinh{\frac{1}{2}\Delta \Gamma_K t + \sin{\phi_K}\sin{\Delta m_K t}}\Big\},
\end{align}
where $\mathcal{A}_2$ denotes the amplitude for the decay $K^0 \to \pi^+ \pi^-$ and $\abs{\mathcal{A}_1}^2/\abs{\mathcal{A}_2}^2 \approx 4.1 \times 10^{-3}$. We define the absolute branching ratio of $K^0(t) \to \pi^- \ell^+ \nu_\ell$ as
\begin{align}
\frac{\int_0^t \Gamma[K^0(t) \to \pi^- \ell^+ \nu_\ell]}{\int_0^t \Gamma[K^0(t) \to \pi^+ \pi^-]}\times 2 \times \mathcal{B}(K^0_S \to \pi^+ \pi^-) \times A(t).
\end{align}
Here, the factor of 2 arises from the difference in the decay widths of $K^0$ and $K^0_S$. $A(t)$ represents the fraction of $K^0_S$ decays that have occurred by time $t$. Considering that at $t = 0.5 \tau_K \approx \tau_{K_S}$, 68\% of the decays have occurred, we define $A(t)$ as
\begin{align}
A(t) \equiv 1 - e^{-(\Gamma_{K_S} + \lambda) t},\label{At}
\end{align}
where $\lambda$ is the correction factor, approximately equal to 0.14$\Gamma_{K_S}$.

\begin{figure}[htbp]
    \centering
    \includegraphics[keepaspectratio,width=6.2cm]{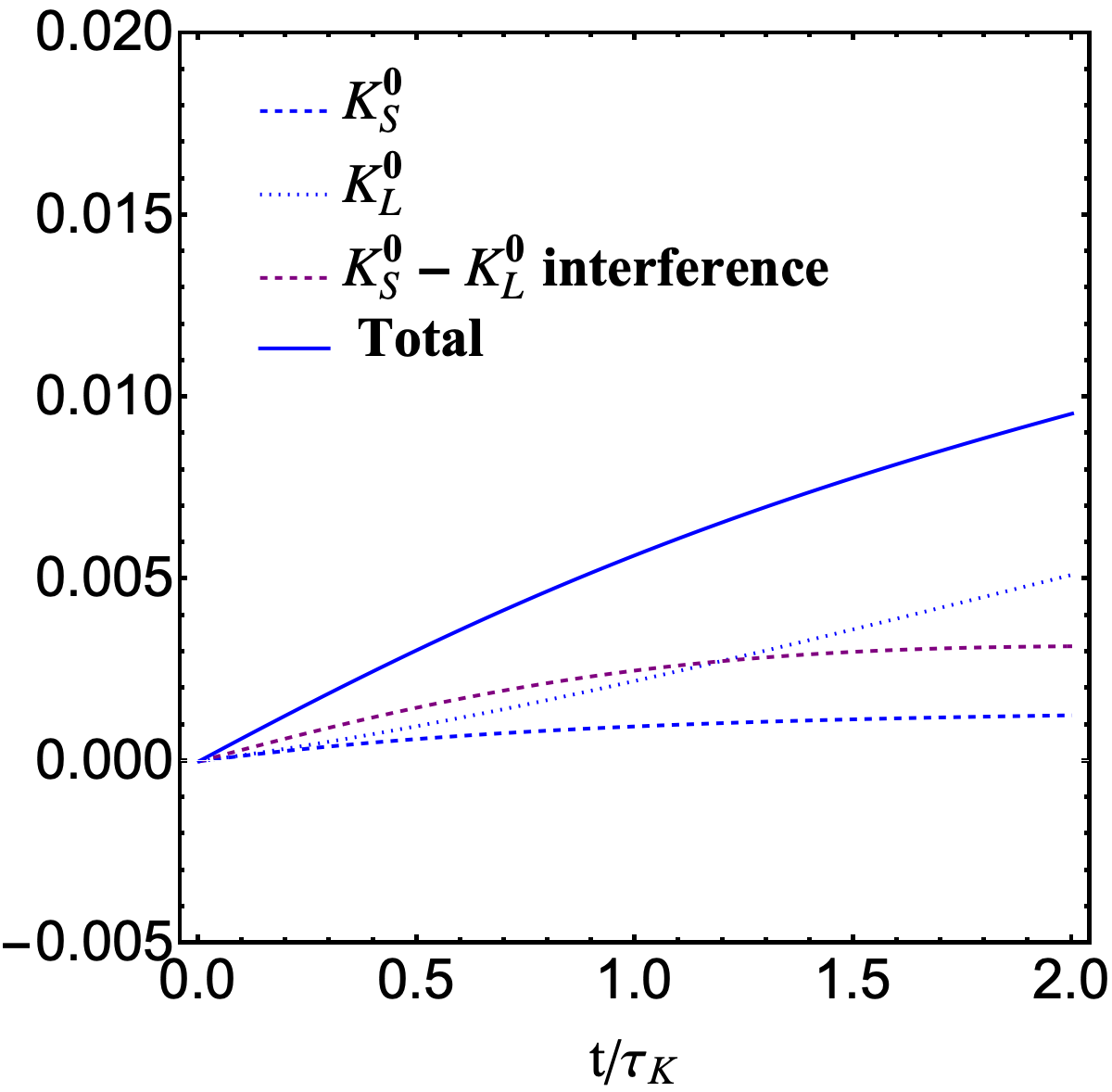}
    \hspace{0.2cm}
    \includegraphics[keepaspectratio,width=5.6cm]{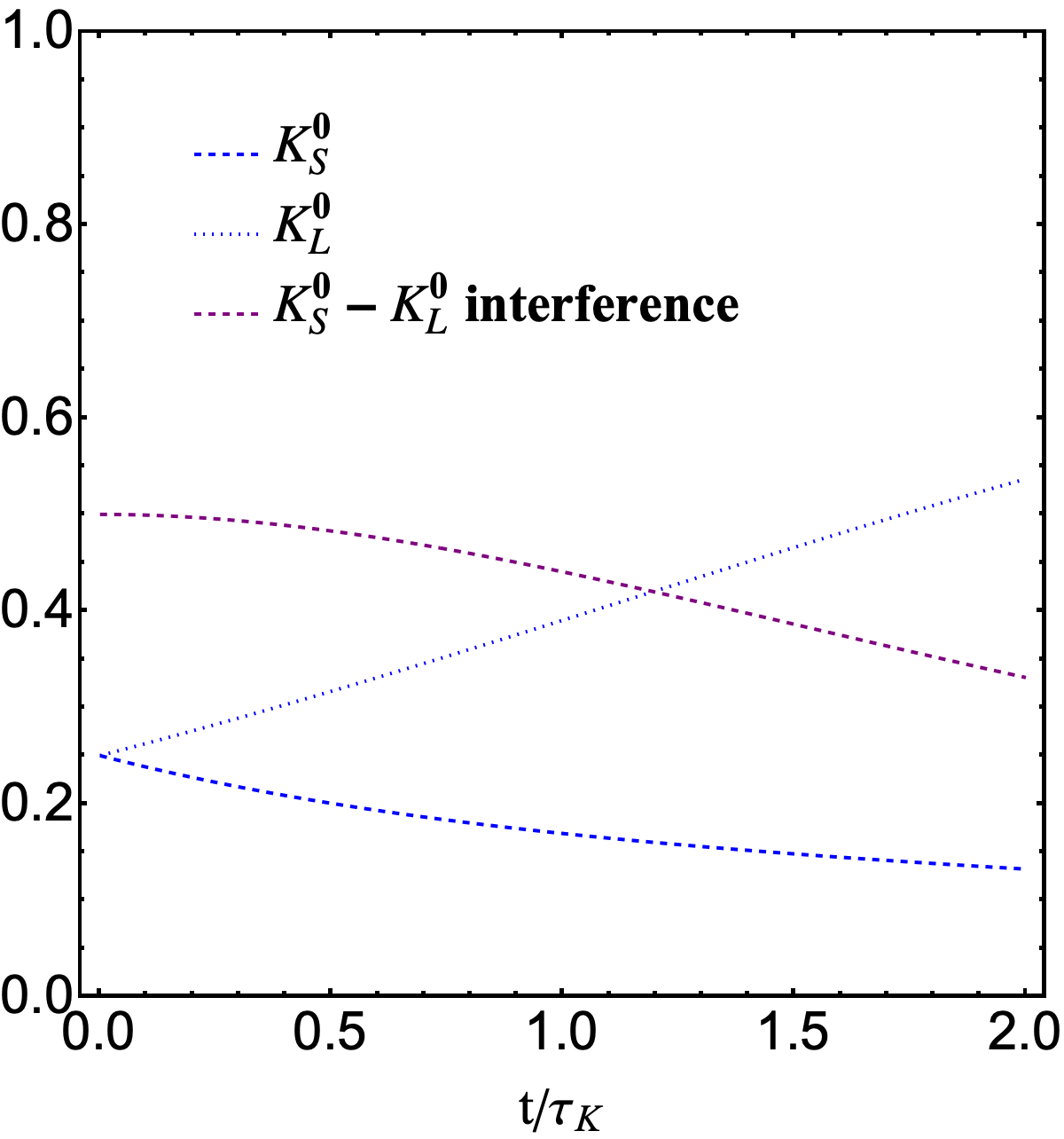}
     \hspace{0.1cm}

    \caption{The left panel displays the time dependence of the absolute branching ratios for the three contributions (with time integrated from 0 to t). The right panel displays the percentage of the total contribution from the three decay widths (with time integrated from 0 to t).}\label{fig:Semileptonic decay}
\end{figure}
\end{appendix}

{The time dependence of the absolute branching ratios of the three contributions to $K^0(t) \to \pi^- \ell^+ \nu_\ell$ is shown in the left panel of Fig.~\ref{fig:Semileptonic decay}. The total branching ratio for $K^0(t) \to \pi^- \ell^+ \nu_\ell$, represented by the blue solid line, increases with the integration time $t$ and reaches 1\% at $t = 2\tau_K$. The branching ratio of $K^0_L$ continuously increases over time and reaches 0.5\% at $t = 2\tau_K$, accounting for half of the total branching ratio. In contrast, the branching ratios associated with $K^0_S$ and $K^0_S - K^0_L$ interference remain nearly constant for $t > \tau_K$. This is because their lifetimes are significantly shorter compared to that of $K^0_L$, with the majority of their decays occurring within $\tau_K$.} The right panel shows the relative contributions of the three terms to the total decay width, integrated from time 0 to $t$. Initially, the $K^0_S - K^0_L$ interference term accounts for 50\%, while the direct decay of $K^0_S$ and $K^0_L$ each contribute 25\%. As the integration time increases from 0 to $2\tau_K$, the interference term decreases to below 35\%, the contribution from $K^0_S$ also diminishes, while the $K^0_L$ contribution increases to over 50\%.

{}


\begin{thebibliography}{}

\bibitem{Sakharov:1967dj}
A.~D.~Sakharov,
Pisma Zh. Eksp. Teor. Fiz. \textbf{5}, 32-35 (1967)

\bibitem{Muller:1960ph}
F.~Muller, R.~W.~Birge, W.~B.~Fowler, R.~H.~Good, W.~Hirsch, R.~P.~Matsen, L.~Oswald, W.~M.~Powell, H.~S.~White and O.~Piccioni,
Phys. Rev. Lett. \textbf{4}, 418-421 (1960)

\bibitem{Christenson:1964fg}
J.~H.~Christenson, J.~W.~Cronin, V.~L.~Fitch and R.~Turlay,
Phys. Rev. Lett. \textbf{13}, 138-140 (1964)

\bibitem{KTeV:1999kad}
A.~Alavi-Harati \textit{et al.} [KTeV],
Phys. Rev. Lett. \textbf{83}, 22-27 (1999)
[arXiv:hep-ex/9905060 [hep-ex]].

\bibitem{BaBar:2001ags}
B.~Aubert \textit{et al.} [BaBar],
Phys. Rev. Lett. \textbf{86}, 2515-2522 (2001)
[arXiv:hep-ex/0102030 [hep-ex]].

\bibitem{Belle:2001zzw}
K.~Abe \textit{et al.} [Belle],
Phys. Rev. Lett. \textbf{87}, 091802 (2001)
[arXiv:hep-ex/0107061 [hep-ex]].

\bibitem{BaBar:2004gyj}
B.~Aubert \textit{et al.} [BaBar],
Phys. Rev. Lett. \textbf{93}, 131801 (2004)
[arXiv:hep-ex/0407057 [hep-ex]].

\bibitem{Belle:2004nch}
Y.~Chao \textit{et al.} [Belle],
Phys. Rev. Lett. \textbf{93}, 191802 (2004)
[arXiv:hep-ex/0408100 [hep-ex]].

\bibitem{LHCb:2013syl}
R.~Aaij \textit{et al.} [LHCb],
Phys. Rev. Lett. \textbf{110}, no.22, 221601 (2013)
[arXiv:1304.6173 [hep-ex]].

\bibitem{LHCb:2019hro}
R.~Aaij \textit{et al.} [LHCb],
Phys. Rev. Lett. \textbf{122}, no.21, 211803 (2019)
[arXiv:1903.08726 [hep-ex]].

\bibitem{Glashow:1970gm}
S.~L.~Glashow, J.~Iliopoulos and L.~Maiani,
Phys. Rev. D \textbf{2}, 1285-1292 (1970)

\bibitem{Grossman:2011zk}
Y.~Grossman and Y.~Nir,
JHEP \textbf{04}, 002 (2012)
[arXiv:1110.3790 [hep-ph]].

\bibitem{Yu:2017oky}
F.~S.~Yu, D.~Wang and H.~n.~Li,
Phys. Rev. Lett. \textbf{119}, no.18, 181802 (2017)
[arXiv:1707.09297 [hep-ph]].

\bibitem{Shen:2023nuw}
Y.~F.~Shen, W.~J.~Song and Q.~Qin,
Phys. Rev. D \textbf{110}, no.3, L031301 (2024)
[arXiv:2301.05848 [hep-ph]].

\bibitem{Song:2024jjn}
W.~J.~Song, Y.~F.~Shen and Q.~Qin,
Eur. Phys. J. C \textbf{84}, no.10, 1030 (2024)
[arXiv:2403.01904 [hep-ph]].

\bibitem{BESIII:2020nme}
M.~Ablikim \textit{et al.} [BESIII],
Chin. Phys. C \textbf{44}, no.4, 040001 (2020)
[arXiv:1912.05983 [hep-ex]].

\bibitem{Belle-II:2018jsg}
E.~Kou \textit{et al.} [Belle-II],
PTEP \textbf{2019}, no.12, 123C01 (2019)
[erratum: PTEP \textbf{2020}, no.2, 029201 (2020)]
[arXiv:1808.10567 [hep-ex]].

\bibitem{Cerri:2018ypt}
A.~Cerri, V.~V.~Gligorov, S.~Malvezzi, J.~Martin Camalich, J.~Zupan, S.~Akar, J.~Alimena, B.~C.~Allanach, W.~Altmannshofer and L.~Anderlini, \textit{et al.}
CERN Yellow Rep. Monogr. \textbf{7}, 867-1158 (2019)
[arXiv:1812.07638 [hep-ph]].

\bibitem{Ai:2024nmn}
X.~Ai, W.~Altmannshofer, P.~Athron, X.~Bai, L.~Calibbi, L.~Cao, Y.~Che, C.~Chen, J.~Y.~Chen and L.~Chen, \textit{et al.}
[arXiv:2412.19743 [hep-ex]].

\bibitem{Charm-TauFactory:2013cnj}
A.~E.~Bondar \textit{et al.} [Charm-Tau Factory],
Phys. Atom. Nucl. \textbf{76}, 1072-1085 (2013)

\bibitem{Achasov:2023gey}
M.~Achasov, X.~C.~Ai, R.~Aliberti, L.~P.~An, Q.~An, X.~Z.~Bai, Y.~Bai, O.~Bakina, A.~Barnyakov and V.~Blinov, \textit{et al.}
Front. Phys. (Beijing) \textbf{19}, no.1, 14701 (2024)
[arXiv:2303.15790 [hep-ex]].

\bibitem{LHCb:2022gnc}
R.~Aaij \textit{et al.} [LHCb],
Phys. Rev. D \textbf{105}, no.9, 092013 (2022)
[arXiv:2202.09106 [hep-ex]].

\bibitem{LHCb:2022cak}
R.~Aaij \textit{et al.} [LHCb],
Phys. Rev. D \textbf{108}, no.5, 052005 (2023)
[arXiv:2208.06512 [hep-ex]].

\bibitem{Golowich:2005pt}
E.~Golowich and A.~A.~Petrov,
Phys. Lett. B \textbf{625}, 53-62 (2005)
[arXiv:hep-ph/0506185 [hep-ph]].

\bibitem{Li:2020xrz}
H.~N.~Li, H.~Umeeda, F.~Xu and F.~S.~Yu,
Phys. Lett. B \textbf{810}, 135802 (2020)
[arXiv:2001.04079 [hep-ph]].

\bibitem{Li:2022jxc}
H.~n.~Li,
Phys. Rev. D \textbf{107}, no.5, 054023 (2023)
[arXiv:2208.14798 [hep-ph]].

\bibitem{Jiang:2017zwr}
H.~Y.~Jiang, F.~S.~Yu, Q.~Qin, H.~n.~Li and C.~D.~L\"u,
Chin. Phys. C \textbf{42}, no.6, 063101 (2018)
[arXiv:1705.07335 [hep-ph]].

\bibitem{Cheng:2024hdo}
H.~Y.~Cheng and C.~W.~Chiang,
Phys. Rev. D \textbf{109}, no.7, 073008 (2024)
[arXiv:2401.06316 [hep-ph]].

\bibitem{ParticleDataGroup:2024cfk}
S.~Navas \textit{et al.} [Particle Data Group],
Phys. Rev. D \textbf{110}, no.3, 030001 (2024)

\bibitem{CLEO:2007rhw}
Q.~He \textit{et al.} [CLEO],
Phys. Rev. Lett. \textbf{100}, 091801 (2008)
[arXiv:0711.1463 [hep-ex]].

\bibitem{Rosner:2006bw}
J.~L.~Rosner,
Phys. Rev. D \textbf{74}, 057502 (2006)
[arXiv:hep-ph/0607346 [hep-ph]].

\bibitem{Li:2012cfa}
H.~n.~Li, C.~D.~Lu and F.~S.~Yu,
Phys. Rev. D \textbf{86}, 036012 (2012)
[arXiv:1203.3120 [hep-ph]].

\bibitem{Qin:2013tje}
Q.~Qin, H.~n.~Li, C.~D.~L\"u and F.~S.~Yu,
Phys. Rev. D \textbf{89}, no.5, 054006 (2014)
[arXiv:1305.7021 [hep-ph]].

\bibitem{Qin:2021tve}
Q.~Qin, C.~Wang, D.~Wang and S.~H.~Zhou,
Front. Phys. (Beijing) \textbf{18}, no.6, 64602 (2023)
[arXiv:2111.14472 [hep-ph]].



\end{thebibliography}
\end{document}